\documentclass[10pt,journal,compsoc]{IEEEtran}

\usepackage{amsmath,amssymb}
\usepackage{multirow}
\usepackage[table]{xcolor}
\usepackage{cite}
\usepackage[pdftex]{graphicx}
\usepackage{hyperref,backref}
\hypersetup{colorlinks=true,citecolor=blue}
\usepackage{url}
\usepackage{array,arydshln}
\usepackage{tabu}
\usepackage{tikz}

\newcommand{\pk}{{\mathnormal{pk}}}
\newcommand{\sk}{{\mathnormal{sk}}}
\newcommand{\F}{{\mathcal{F}}}
\newcommand{\Z}{{\mathbb{Z}}}
\newcommand{\G}{{\mathbb{G}}}
\newcommand{\N}{{\mathbb{N}}}

\newcommand{\M}{{\mathcal{M}}}

\newcommand{\enc}{{\mathsf{E}}}
\newcommand{\re}{{\mathsf{Re}}}
\newcommand{\kg}{{\mathsf{Kg}}}
\newcommand{\dec}{{\mathsf{D}}}
\newcommand{\tkg}{{\mathsf{TKg}}}
\newcommand{\tdec}{{\mathsf{TD}}}
\newcommand{\pke}{{\mathcal{E}}}
\newcommand{\PT}{{\mathcal{M}_\pk}}
\newcommand{\CT}{{\mathcal{C}_\pk}}
\newcommand{\RT}{{\mathcal{R}_\pk}}
\newcommand{\A}{{\mathcal{A}}}

\newcommand{\spara}{{\lambda}}

\newcommand{\state}{{\mathsf{s}}}
\newcommand{\mexp}{{\mathsf{exp}}}
\newcommand{\mmul}{{\mathsf{mul}}}
\newcommand{\usr}{{\mathnormal{u}}}
\newcommand{\qry}{{\mathsf{q}}}
\newcommand{\ev}{\bar{\mathsf{v}}}
\newcommand{\share}{{\mathsf{Sh}}}
\newcommand{\open}{{\mathsf{Rc}}}

\renewcommand{\Pr}{{\mathbb{P}}}
\newcommand{\negl}{{\mathsf{negl}}}
\newcommand{\vecq}{{\mathbf{q}}}
\newcommand{\bad}{{\mathsf{MalGrp}}}
\newcommand{\gp}{{\mathcal{G}}}

\newcommand{\ct}[1]{{\bar{#1}}}
\newcommand{\rct}[1]{{\hat{#1}}}
\newcommand{\bs}[1]{{\boldsymbol{#1}}}
\newcommand{\popk}{{\mathsf{zk\text{-}PK}}}

\newcommand{\pws}{{\mathsf{PWS}}}
\newcommand{\vs}{{\mathsf{zk\text{-}CS}}}
\newcommand{\podl}{{\mathsf{zk\text{-}DL}}}

\newcommand{\Sim}{{\mathcal{S}}}
\newcommand{\lang}{{\mathcal{L}}}

\newcommand{\real}{{\textsc{real}}}
\newcommand{\ideal}{{\textsc{ideal}}}

\newtheorem{thm}{Theorem}{\bfseries}{\itshape}
{\bfseries}{\itshape}
\newtheorem{lem}{Lemma}{\bfseries}{\itshape}
{\bfseries}{\itshape}
{\bfseries}{\itshape}
\newtheorem{df}{Definition}{\bfseries}{\itshape}
{\bfseries}{\itshape}

{\itshape}
{\itshape}
{\normalfont}

\begin{document}

\title{Private Web Search with an Expected Constant Round} 
 
\author{Myungsun Kim %
\IEEEcompsocitemizethanks{
\IEEEcompsocthanksitem M. Kim is with the Department of Information Security, 
The University of Suwon, Hwaseong-si, Gyeonggi-do 18323, South Korea. %
E-mail: msunkim@suwon.ac.kr}%
}

\IEEEtitleabstractindextext{
\begin{abstract}

Web searching is becoming an essential activity 
because it is often the most effective and convenient  way of finding information. 
However, a Web search can be a threat to the privacy of the searcher
because the queries may reveal sensitive information about the searcher. 
Private Web search (PWS) solutions allow users to find information on the Internet while preserving their privacy. 
Here, privacy means maintaining the confidentiality of the identity of the communicating users. 
According to their underlying technology, existing PWS solutions can be divided into three types: 
proxy-based solutions, obfuscation-based solutions, and cryptography-based solutions. 
Of these, cryptography-based PWS (CB-PWS) systems are particularly interesting 
because they provide strong privacy guarantees in the cryptographic sense. 
In this paper, we present a round-efficient CB-PWS protocol 
that preserves computational efficiency compared to other known CB-PWS systems. 
Assuming a broadcast channel, our protocol is a \emph{four-round} cryptographic  scheme that requires $O(n)$ communication complexity. 
However, if only point-to-point interaction is available, with the users emulating the broadcast channel,
our protocol requires an expected $O(1)$-round complexity and the same computation and communication overhead.
Further analyzing the efficiency of our protocol shows 
that  our proposal requires only $3n$ modular exponentiations for $n$ users. 
To evaluate  the security of our protocol, we  demonstrate that  our construction is secure in terms of a semi-honest model. 
We then discuss how to enhance its security to render it secure in the presence of malicious adversaries.
We provide a specific protocol for managing users' groups, which is also an advantage over existing systems.
\end{abstract} 

}

\maketitle


\IEEEraisesectionheading{%
\section{Introduction}\label{sec:introduction}%
}

\IEEEPARstart{U}sing a private Web search (PWS) prevents Web-search service providers such as Google and Bing  from building user
profiles while still allowing the users to enjoy full search functionality when performing Web searches.
User profiling is usually defined as the process of implicitly developing a user profile 
from the search-engine queries submitted by the user. 
The service provider can then use user-profiling information, which might include the users private interests and preferences,
to assign the user to a predefined user class, such as a demographic or taste category, 
or to capture the online behavior of the user. 
Although user profiles may enable service providers to offer a better service,
clearly this raises \emph{privacy concerns} because sensitive information,
such as a user's name and location, can be inferred from search-engine queries. 
In addition to the query terms themselves, other information, such as the source IP address 
and timestamp, may reveal  sensitive information about the user.

Various approaches have been proposed to address this problem. 
Balsa et al.~\cite{BTD12} identified three types of  PWS protocols, in terms of the key techniques used for anonymity. 
First, if a PWS solution introduces a proxy server to submit  
query words on behalf of the user, it is  called a  proxy-based technique (e.g., ~\cite{Ano,Scr,SJBF07}).  
A second group of PWS solutions enables users to submit a collection of queries in such a way 
that the real query term is buried among many other (fake) terms, thereby obscuring the identity of the real term. 
This is called obfuscation-based PWS (e.g.,~\cite{ESM06,DSC09,RF10}). 
The third approach relies on cryptographic tools to eliminate the possibility of linking 
users to their query terms. This is called  cryptography-based PWS (CB-PWS) 
(e.g., ~\cite{CVH09,LW10,RCV11,KK12}) and is used in our proposal.

In this paper, we are particularly interested in constructing an efficient CP-PWS protocol. 
Our choice relates to our technical standpoint 
on the trade-off between performance and security. 
Controversially, PWS solutions based on proxy or obfuscation 
techniques place more importance on performance than on security. 
Therefore, we can say that their goal is to find a way of enhancing security 
while not affecting the performance of legacy Web-searching services. 
On the other hand, CB-PWS solutions pursue strong security 
in spite of the high possibility of degrading search performance. 
Accordingly, PWS protocols of this type aim to find methods  that  minimize the performance degradation.

Our goal in this work is twofold.

First, we construct an efficient CB-PWS protocol with \emph{constant-round} complexity,  
but with the same computation and communication costs as a standard protocol. When designing an interactive protocol, 
researchers  will always investigate the round complexity of their method,
because interaction over a network is 
usually affected by lagging or network congestion, becoming the most time-consuming aspect  of the operation.  
We have many good examples of this type of secure protocol (e.g.,~\cite{BB89,BMR90,CCKM00,IK00}). 
Unfortunately, although CB-PWS solutions are typical examples of interaction-intensive protocols,  
round complexity has not been taken into account as a key efficiency metric to accompany computation 
and communication complexity. Therefore, it is very important to devise protocols 
that require a minimal number of rounds to complete. 
Our challenging goal is to find a CB-PWS solution with $O(1)$-round complexity for a number of users.

Second,   we provide our efficient CB-PWS protocol with \emph{simulation-based security},
focusing on the model of \emph{malicious users}.
We have observed that existing CB-PWS systems  have not undergone rigorous proof of security, 
although ad-hoc security analysis has been performed in some cases.
The only exception is  Lindell and Waisbard's scheme~\cite{LW10}, but this has involved only game-based security.
We believe that it important to define precisely, and prove formally, the security 
that a cryptographic protocol offers,
because the  history of cryptographic-scheme design 
gives good reasons to distrust heuristic approaches.

Having stated the technical goals of this work and
before giving a conceptual description of our PWS solution, 
we now summarize our contributions and present a high-level overview of our construction. 
For readers who would first like to review existing work in the PWS field 
to better understand where our work fits in, 
Section~\ref{sec:related-work} surveys the PWS state of the art.
However, in explaining our work, we will sometimes  
also give  a brief survey of what is known in the literature about the relevant PWS issues.

\subsection{Our Contributions}\label{subsec:our-contr}

The technical contributions of this paper can be summarized as:

\begin{itemize}

\item We design a constant-round CB-PWS solution with comparable computation and communication complexity to existing systems, 
assuming the existence of a broadcast channel. Without the broadcast, our protocol expects to involve constant 
rounds for a number of users while preserving the computation and communication cost.

Kang et al.~\cite{KGK15} argued that their CB-PWS protocol has a constant number of rounds to complete. 
However, their work does not consider a network where only point-to-point communication is allowed.
Furthermore, their scheme does not provide a specific protocol to set up a group of users, 
which should be a mandatory part of any CB-PWS protocol. 
In fact, no CB-PWS solutions other than that of Lindell and Waisbard   have handled this issue. 
These CB-PWS schemes do not provide a way to determine a group leader among $n$ users, 
but it is well known that group-leader election is a nontrivial issue, 
particularly with the constant-round restriction.  
Therefore, it is not clear if Kang et al.'s scheme will complete the protocol activities in a constant number of rounds.

Kim and Kim~\cite{KK12}  proposed a round-efficient CB-PWS scheme, 
but it significantly restricts the message-space size (at most $5$ bits for a group of 20 users).  
Therefore, it does not lead to a preferable solution to the problem.

\item We provide a so-called formal proof of security, 
which uses a simulation-based technique. 
It is clearly preferable to prove security using a standard simulation-based technique with a universally composable model. 
Differently from~\cite{LW10},  we define an ideal functionality to fit the PWS model and show that our protocol 
can be transformed efficiently into the ideal functionality.

\item We provide a specific protocol for efficiently creating a group of $n$ users. 
In addition, we solve the inefficiency problem in Kang et al.'s work. 
Even if Kang et al.  resolved the incompleteness issues above, their scheme is far from 
efficient in the sense of computation complexity. 
This is because the group manager has $O(n^3\log^2n)$ computation complexity in the number of users $n$. 
Therefore, their scheme can run efficiently only for a very small number of users. ($3\sim 4$). 
We develop a solution to these technically crucial problems  
without restricting the plaintext length, which indeed is the primary difference from Kang et al.'s results.

\end{itemize}

\subsection{A Key Idea behind Our Scheme}\label{subsec:overview}

We now describe  the properties of our scheme,
compared to existing CB-PWS solutions, from a design-philosophy viewpoint.

Techniques common to our protocol and existing protocols are, first, to encrypt 
users' query terms via a proper encryption algorithm, and then 
to rerandomize and mix the ciphertexts to remove linkability between
users and their query terms. Figure~\ref{fig:old-design} shows 
these general protocol actions, together with the number of rounds required to 
complete subprotocols such as a shuffle protocol.

\begin{figure}[h!]
\centering
\begin{tabular}{ccc} \hline
\multirow{2}{*}{Protocol flow} & Main & \# of\\
				& operations & Rounds \\ \hline
$
\begin{array}{ccc}
u_1 &\cdots & u_n\\
  \downarrow &  & \downarrow \\
  \ct{\qry}_1=\enc(\qry_1) & \cdots & \ct{\qry}_n=\enc(\qry_n)
\end{array}
$ & Encrypt & 1\\
$\Downarrow$ & & \\
\framebox[1.2\width]{Shuffle } & Remask \& & \multirow{2}{*}{$\boldsymbol{n}$}\\
$(\pi,r_1,\cdots,r_n)$ & Mix & \\
$\Downarrow$ & & \\
$\{\rct{\qry}_1,  \ldots, \rct{\qry}_n\}$ & &\\
$\Downarrow$ & & \\
\framebox[1.1\width]{Group decryption} & Decrypt& $1$\\
$\Downarrow$ & & \\
$\{\qry_1, \ldots, \qry_n\}$ & &\\
$\Downarrow$ & & \\
\framebox[1.1\width]{Web query} & & $1$\\
$\Downarrow$ & & \\
\multirow{2}{*}{$\{a_1,\ldots,a_n\}$}  & Broadcast  & \multirow{2}{*}{1}\\ 
& query result & \\ \hline
\end{tabular}
\caption{General CB-PWS design framework}\label{fig:old-design}
\end{figure}

\smallskip
\noindent\textbf{Existing works and their basic framework}. %
We are mainly interested in \emph{Shuffle} phase because 
this step requires $O(n)$ rounds for a number of users $n$.
In existing solutions, each user  $u_i$ first computes an encryption of its own query term $\qry_i$, 
denoted by $\ct{\qry}_i$, under an encryption algorithm $\enc$.
Then all users  join and run a shuffle protocol,  taking
as inputs a private permutation $\pi$ over the set $\{1,2,\ldots,n\}$ and 
a set of fresh randomizers $(r_1,r_2,\ldots,r_n)$.
This shuffle protocol should be performed in a relay manner among all users
because  the sequential shuffles they employ can only achieve unlinkability
in this manner. Therefore, a round complexity of $O(n)$ seems to be unavoidable.

\smallskip
\noindent\textbf{Our idea}.  We utilize a similar design framework.
However, our scheme does not require the users to engage in a sequential shuffle protocol. However, note that this does not mean that
our scheme does not use the shuffle protocol at all. 
Similarly, our scheme also  outputs a list of rerandomized and  mixed ciphertexts during its execution,
but without any interaction with other users, i.e., in a stand-alone mode.
To achieve this, instead of using a query term $\qry_i$, we encrypt a share of $\qry_i$ 
after letting its other shares be distributed to other users.
This is the main difference between our protocol and other existing solutions.
More specifically, consider a situation before invoking the shuffle protocol in both cases.
In existing solutions, all users have the same list of ciphertexts 
$\{\ct{\qry}_1,\ct{\qry}_2,\ldots,\ct{\qry}_n\}$
in clear. 
In contrast, users in our solution have a different list of ciphertexts at this point, i.e.,
$\{\ct{\qry}_{i,1},\ct{\qry}_{i,2},\ldots,\ct{\qry}_{i,n}\}$
by denoting the $j$-th share of $\qry_i$ as $\qry_{i,j}(1\leq j\leq n)$. 
Following this basic technical property,
our protocol   does not require 
any interaction between neighbors and  does not incur 
a number of rounds proportional to the number of users.

However, we do face a new problem. 
Because all users' query terms have been distributed in the form of shares and then encrypted,
after performing group decryption, 
all shares should be presented to an algorithm that can  reconstruct 
a mixed list of original query terms $\{\qry_{\pi(1)},\ldots,\qry_{\pi(n)}\}$.
Because we cannot match the decrypted shares to their original messages,
a na\"{i}ve method involving trial and error runs the reconstruction algorithm $O(n^2)$ times.
This incurs $O(n^3\log^2n)$ computational complexity in total,
assuming that fast interpolation can be done by $O(n\log^2n)$ multiplications.
In this work, we  develop a lightweight solution to resolve this computational 
problem at the cost of some slight damage in the plaintext domain.

\smallskip
\noindent\textbf{The outline}. The structure of this paper is as
follows. Section~\ref{sec:background} introduces basic definitions and cryptographic 
primitives: secret sharing and public-key homomorphic 
encryption. The system model for running our scheme is 
described in Section~\ref{sec:model}. Section~\ref{sec:pws} provides a detailed description
of our construction, together with a full description of the
performance and  security analysis of our protocol. 
Section \ref{sec:group} deals with a key subprotocol by which 
a group can be constructed and a group manager can be elected.
Section~\ref{sec:related-work} contains a review of the relevant literature and we make some concluding remarks in Section~\ref{sec:conc}.

\section{Definitions and Basics}\label{sec:background}

In this section, we review briefly the concepts and notations in cryptographic building blocks. 
We then  give a definition of security for CB-PWS which will be used in proving formally 
that  our proposal is secure in the presence of malicious adversaries.

\smallskip
\noindent\textbf{Mathematical notation}. %
If $A$ is a probabilistic polynomial-time (PPT) machine, we use $a\gets A$ 
to denote making $A$ produce an output according to its internal randomness.
In particular, if $U$ is a set, then $r\xleftarrow{\$} U$ is used to denote sampling  
from the uniform distribution on $U$. 
Letting $U^{t}[x]$ be the set of all polynomials of degree
$0,\ldots,t$ with coefficients from  $U$,
we denote by $f\xleftarrow{\$}U^{t}[x]$ a polynomial chosen independently and uniformly from
$U^{t}[x]$.

A \emph{negligible} function, denoted  by $\negl(\spara)$, is a $\nu(\spara)$ 
such that $\nu(\spara) = o(\spara^{-\kappa})$ for every fixed constant $\kappa$. 
For $n\in\mathbb{N}$, $[n]$ denotes the set $\{1,\ldots,n\}$.
For any integer $x$ the length of the binary representation of $x$ is denoted by $|x|$, 
but when the context is clear we  also use $|X|$ to denote the cardinality of  a set $X$ .

Let $X=\{X_\spara\}_{\spara\in\N}$ and $Y=\{Y_\spara\}_{\spara\in\N}$ be ensembles.
Two ensembles $X$ and $Y$ are computationally indistinguishable,
denoted by $X\overset{\text{c}}{=} Y$, if 
for every PPT algorithm $D$ and 
all $\spara\in\N$, 
\begin{equation*}
\left|\Pr[D(X_\spara,1^\spara)=1]-\Pr[D(Y_\spara,1^\spara)=1]\right| < \negl(\spara)
\end{equation*}

\subsection{Threshold Secret Sharing}\label{subsec:ss}

We assume that the readers are familiar with the notion of secret sharing 
and Shamir's implementation.

A secret sharing scheme (e.g.,~\cite{Sha79,Bla79}) consists typically of an algorithm 
for sharing a secret  and an algorithm  for reconstructing the shared secret. 
We denote the first as $\share$ and the second as $\open$. 
Shamir's scheme is based on polynomial interpolation and involves $n$ points on the Cartesian plane. 
Using  these $n$ points, a unique polynomial $f(x)$ over a finite field  is guaranteed to exist 
such that $f(x) = y$ for each of the points given. 
For a concrete instantiation of Shamir's scheme, we need first 
to determine an appropriate field $F$ for the subsequent
modular arithmetic. For instance, we can take $F$ as $\Z_{\tilde{q}}$ 
for a prime $\tilde{q}$. Given $f\xleftarrow{\$}{\Z_{\tilde{q}}}^{n-1}[x]$ and a secret $\qry\in \Z_{\tilde{q}}$, 
we use $f$ whose constant term  has been replaced by the $\qry$, 
in running algorithms $\share$ and $\open$ subsequently.

However, Shamir's secret sharing is insufficient for settings  involving active corruption, 
where an adversary may corrupt users in an arbitrary way. 
In particular, ordinary secret sharing is not effective  
in  settings that may require a secret to be secured for a long period of time.
Indeed, standard secret sharing schemes are no longer secure after some number $t$ of users have been corrupted. 
To avoid advanced threats which, given sufficient time, 
will successfully corrupt sufficient users to break the threshold that guarantees security, 
a proactive security model can be considered, such as Ostrovsky and Yung~\cite{OY91}.
Alternatively, verifiable secret sharing (VSS) could be considered, to 
prevent a malicious dealer from distributing spurious shares.

Fortunately, there are three considerations that free us from imposing
a heavy computation burden on the users to make up for the  security weakness
 of ordinary secret sharing in a stronger security model:
\begin{enumerate}
\item Because  it is not necessary to preserve the confidentiality 
of users' query terms for a long period, we do not need to consider proactive security.

\item Users distribute encryptions of shares of their own query words 
	rather than the shares themselves.
	The more important thing is that the shares should be
	encoded into a specific form before being given to an encryption algorithm.
	Whether the decryption of encrypted share is well-formed can be checked later.
	If the decryption is not a well-formed share, then the decryption is simply discarded.
 
\item Our technique for encoding secret shares involves both additions and multiplications,
	but, because our underlying encryption has only a group homomorphism, 
	we cannot plug our encoding scheme  into an existing VSS scheme.
\end{enumerate}

\subsection{Threshold Homomorphic Encryption}\label{subsec:pkc}

A  public-key  encryption scheme $\pke=(\kg,\enc,\dec)$ comprises the following algorithms:
\begin{itemize}

\item $\kg$ is a randomized algorithm that takes a security parameter $\lambda$ 
as input and outputs a secret key $\sk$ and
a public key $\pk$. $\pk$ defines a plaintext space $\PT$ and a ciphertext space $\CT$.

\item $\enc$ is a randomized algorithm that takes $\pk$ and a
  plaintext $m\in\PT$ as input and outputs a ciphertext
  $c\in\CT$. Note that this process is usually randomized using a randomizer $r\in\RT,$
  denoted by $c = \enc_\pk(m;r)$.

\item $\dec$ takes $\sk$ and $c\in\CT$ as input and outputs the plaintext $m$.

\end{itemize}
We say that an encryption scheme is \emph{correct} if, 
for any $(\pk,\sk) \xleftarrow{} \kg(1^\lambda)$ and any $m\in\PT$,
 $m=\dec_\sk(\enc_\pk(m))$.

We say that a public-key cryptosystem
$\pke$ is \emph{homomorphic} for the binary relations $(\oplus,\otimes)$ if for all $(\pk,\sk)\gets\kg(1^\spara)$, 
given $\PT$ and $\CT$, $(\PT,\oplus)$ forms a group and  $(\CT,\otimes)$ forms a group. 
Further, for all $c_1,c_2\in\CT$, $\dec_\sk(c_1\otimes c_2)=\dec_\sk(c_1)\oplus\dec_\sk(c_2).$

Informally,  when two ciphertexts are combined in a
specific manner,  the resulting ciphertext encodes the combination of 
the underlying plaintexts under a specific group operation, usually multiplication or addition.
As a consequence, a cryptosystem's homomorphic property allows us to perform \emph{rerandomization}: 
given a ciphertext $c$, anyone can create a different ciphertext $\bar{c}$ 
that encodes the same plaintext as $c$. 
Therefore, given a group homomorphic cryptosystem $\pke$, 
we can define the rerandomization algorithm as follows:
\begin{equation*}
\re_\pk(c;\gamma):=c\otimes\enc_\pk(0;\gamma),
\end{equation*}
where $0$ is an  identity such that $\forall m\in\PT,
m\oplus 0=m$.

\smallskip
\noindent\textbf{Security  for  homomorphic encryption}. %
Because group homomorphic encryption (GHE) allows malleability in ciphertexts, 
we need only discuss security for homomorphic cryptosystems, the  
so-called \emph{semantic security}.

Semantic security was first defined by Goldwasser and Micali~\cite{GM84}. 
Intuitively, a cryptosystem is said to be semantically secure if, given a ciphertext $c$, 
an adversary cannot determine any property of the underlying plaintext $m$. Specifically, 
an adversary cannot extract any semantic information about plaintext $m$ from an encryption of $m$.

We say that a GHE scheme $\pke=(\kg,\enc,\dec)$ is \emph{semantically secure} 
if, for all polynomial-time algorithms $\A=(\A_1,\A_2)$,
\begin{equation*}
\underset{b,r}{\Pr}\left[ \left.\begin{array}{l}
			(\pk,\sk)\gets\kg(1^\spara);\\
			(m_0,m_1,\state)\gets \A_1(\pk);\\
			c_b\gets\enc_\pk(m_{b};r);\\
			b'\gets\A_2(m_0,m_1,c_b,\state)
			\end{array}\right|b=b'\right]-\frac{1}{2}\leq\negl(\spara),
\end{equation*}
where $b,b'\in\{0,1\}$ and $\state$ is the state information of $\A$.

\smallskip
\noindent\textbf{Threshold GHE.} %
We require a threshold group-homomorphic encryption scheme.
This property is satisfied by most known homomorphic encryption schemes, including 
Goldwasser-Micali~\cite{GM84}, El Gamal~\cite{ElG85}, Paillier~\cite{Pai99} and threshold Paillier~\cite{FPS00}.
As mentioned above, our scheme does not require a particular type of GHE scheme. 
However, because El Gamal's scheme is 
somewhat more efficient than Paillier's scheme in the computational sense, 
we review briefly the El Gamal encryption scheme and its threshold variant.

For large primes $p$ and $q$   such that $q|(p-1)$, let $\G_q$ be 
the unique subgroup of $\Z_p^\times$ of order $q$, and let $g$ be a generator of $\G_q$.
Because any element $1\neq \beta\in\G_q$ generates the group, the discrete logarithm of $\alpha\in\G_q$ 
with respect to the base $\beta$ is defined as usual. 
All computations in the remainder of this paper are modulo $p$ unless otherwise noted.

The standard El Gamal encryption scheme is as follows:
\begin{itemize}
\item  $\kg(1^\lambda)$ outputs a group description $(\G_q,g,p,q)$ by taking the security parameter $\lambda$, 
then publicly opening $y=g^x$ and keeping $x$ secret, where $x\xleftarrow{\$} \Z_q^\times$.
\item $\enc_\pk(m;r)$ outputs $c=(g^r,m\cdot y^r)$ with $r\xleftarrow{\$}\Z_q^\times$.
\item $\dec_\sk(c)$ first parses $c$ into $(\alpha,\beta)$ and outputs $\beta\cdot\alpha^{-x}$.
\end{itemize}
We can easily verify that the El Gamal encryption scheme is multiplicatively homomorphic. 
Its rerandomization algorithm $\re_\pk(c;\gamma):=c\otimes\enc_\pk(1_\G;\gamma)$ is given by
\begin{equation*}
c\otimes\enc_\pk(1_\G;\gamma)=(\alpha\cdot g^\gamma,\beta \cdot 1_\G\cdot y^\gamma)=(g^{r+\gamma},m\cdot y^{r+\gamma}),
\end{equation*}
where the $\otimes$ operation means componentwise group multiplication
and $1_\G$ is the identity element in $\G_q$. 

\smallskip
\noindent\textit{Distributed key generation.} %
Each participant chooses $x_i\xleftarrow{\text{\$}}\Z_q^\times$ and publishes 
$y_i=g^{x_i}$. The public key is $y=\prod_{i=1}^Ny_i$, and
the secret key is $x=\sum_{i=1}^Nx_i$. This requires $N$ multiplications,
but their computational cost is negligible compared to exponentiations.
Further broadcast round complexity is $O(1)$.
We write this algorithm as $\tkg(1^\spara,N)$.

\smallskip
\noindent\textit{Distributed decryption.} %
Given an encryption $c=(\alpha,\beta)$,
each participant publishes their decryption share $\alpha^{x_i}$. 
The plaintext can be derived by computing
$\frac{\beta}{\prod_{i=1}^N\alpha^{x_i}}$. As in key generation, 
decryption can be performed in a constant number of rounds, requiring one exponentiation.
We write this algorithm as $\tdec_{x_i}(c)$.

\subsection{Message Space Compatibility}\label{subsec:compt}

As described briefly in \S\ref{subsec:overview}, our protocol requires 
that each share of a query be encrypted by a threshold GHE scheme. 
Accordingly, all shares from the $\share$ algorithm of a secret sharing scheme 
need to be in the plaintext domain of the  encryption scheme.

Let $F$ be an underlying field for Shamir's sharing scheme
and let $\M$ be the plaintext domain of the GHE scheme.
To resolve this message  compatibility issue, 
$F$ needs to be embedded into $\M$. 
Specifically, our protocol requires that $|\M|\geq |F|+2\lceil\log n\rceil,$  
for the number of users $n$.
In particular, letting $F=\Z_{\tilde{q}}$ and $\M=\G_q$ for two primes $\tilde{q}$ and $q$,
we will take $\tilde{q}$ and $q$ such that 
$\lceil\log q\rceil= \lceil\log \tilde{q}\rceil+2\lceil\log n\rceil$. 
We will discuss a technical reason for these parameter selections in Section~\ref{sec:pws}.

\subsection{Security Definition}\label{subsec:sec-def}

We now present a security definition for our CB-PWS scheme.  
In principle, we follow the standard definition for secure multiparty computation~\cite[\S7]{Gold04}.

\smallskip
\noindent\textbf{Simulation-based security.} %
We define the ideal execution of a function $Q$ on inputs $(\qry_1,\ldots,\qry_n)$
and security parameter $\spara$ as the outputs of the honest users and 
the adversary $\A$ from the above ideal execution. 
More specifically, let $Q:(\{0,1\}^*)^n\rightarrow(\{0,1\}^*)^n$ be 
an $n$-ary functionality and let $Q_i(\qry_1,\ldots,\qry_n)$ be the $i$-th element
of $Q(\qry_1,\ldots,\qry_n)$.
For an index set $J=\{i_1,\ldots,i_t\}\subset[n]$ such that $t<n$, 
let $I=[n]\backslash J$  and 
denote by $Q_J(\qry_1,\ldots,\qry_n)$ the sequence $Q_{i_1}(\qry_1,\ldots,\qry_n),$ $\ldots,$
$Q_{i_t}(\qry_1,\ldots,\qry_n)$.
A pair $(J,\Sim)$ where $\Sim$ is a PPT algorithm, 
represents an ideal-model adversary. 
The ideal execution of $Q$ under $(J,\Sim)$ on input $\vecq=(\qry_1,\ldots,\qry_n)$, 
denoted by $\ideal_{Q,\Sim(\cdot),J}(\vecq,\spara)$,
is defined as the output pair of the honest users and the ideal-model adversary $\Sim$ 
from the ideal execution.
The output of the honest users is $Q_{I}(\vecq')$ and the input of $\Sim$ is 
$(\qry_{j\in J},J,Q_J(\vecq'))$, where $\vecq':=(\qry_1',\ldots,\qry_n')$ 
such that $\qry_i'$ is given by $\Sim$ for $i\in J$ and $\qry_i'=\qry_i$ otherwise.

Let $\Pi$ be an $n$-party protocol for computing $Q$, defined as above.
In the real model, there is no trusted party and the users communicate directly 
with each other. The real-model adversary $\A$ controls the corrupted users and 
therefore sends all messages of its choice in their place.
Further, the adversary is not obliged to follow the specifications 
prescribed in the protocol $\Pi$. The real execution of $\Pi$ on
inputs  $\vecq$ and security parameter $\spara$,
denoted by $\real_{\Pi,A(\cdot),J}(\vecq,\spara)$,
is then defined as the output vector of the honest users and the real adversary $\A$
from the real execution of $\Pi$.

Following the ideal-vs-real standard simulation technique, 
we also  require that
a secure protocol emulates the ideal execution in the real model 
where a trusted party does not exist. 
We consider only a \emph{static} adversary, which is not allowed to corrupt
a user during protocol execution. Implicitly, we assume that the adversary 
takes as input some auxiliary information.

We can now describe the security situation more formally.

\begin{df}[Security in the malicious model]\label{df:sec}
Let $Q$ and $\Pi$ be defined as above.
Protocol $\Pi$ is said to securely compute $Q$ 
if, for every PPT algorithm $\A$, there exists a PPT algorithm $\Sim$, such that for every $J\subset[n]$,
\begin{equation*}
\left\{\ideal_{Q,\Sim,J}(\vecq,\spara)\right\}_{\vecq,\spara}\overset{\text{c}}{=}\left\{\real_{\Pi,\A,J}(\vecq,\spara)\right\}_{\vecq,\spara}
\end{equation*}
\end{df}

\smallskip
\noindent\textbf{Hybrid execution.} %
In our construction, we will use  zero-knowledge proof (ZKP) protocols as subroutines.
A standard technique for plugging a secure  subprotocol $\Pi_\mathcal{G}$ computing a functionality $\mathcal{G}$ 
into a protocol $\Pi$ is to adopt a hybrid model. 
In this model, any execution of the protocol $\Pi$ calling $\Pi_\mathcal{G}$ as a subprotocol 
requires that users interact with each other as in the real model, but access  
the ideal functionality of $\mathcal{G}$ as in the ideal model.
Specifically, when user $\usr_i$ needs to send a message $\qry_i$ to the trusted party,
it begins to execute $\Pi_\mathcal{G}$ on input $\qry_i$ instead.
On the other hand, when the execution of $\Pi_\mathcal{G}$ ends with output $a_i$,
user $\usr_i$ continues with $\Pi$ as if $a_i$ were given by the trusted party.
Then, by the composition theorem of~\cite{Can00}, if $\Pi_\mathcal{G}$ securely computes $\mathcal{G}$,
then the output distribution of a protocol $\Pi$ in a hybrid execution with $\mathcal{G}$ is 
 computationally indistinguishable from the output distribution of $\Pi$ invoking $\Pi_\mathcal{G}$.
Therefore, we need only analyze the security of $\Pi$ 
when working with $\mathcal{G}$ in a hybrid model.

\smallskip
\noindent\textbf{The  ideal functionality of PWS.} %
We now describe an ideal functionality of PWS, where 
each user's input is a search keyword $\qry_i$.
Given a query $\qry_i$, when its query result corresponds to $a_i$,
 the functionality outputs  the union of all $a_i$ results to all users.
If no results are found, $u_i$ is given a null string $\perp$. More formally:
\begin{df}\label{df:ftn}
Let $\qry_i$ be a query word of user $u_i$ and $a_i$ be a corresponding query result
(without loss of generality, we assume that all query words have the same size). Then the ideal functionality $\F_{\pws}$ is:
\begin{equation*}
(\qry_1,\ldots,\qry_n)\mapsto (\{ a_1,\ldots,a_n\},\ldots,\{ a_1,\ldots,a_n\})
\end{equation*}
\end{df}

In the following sections,  we present details of our PWS construction for 
a protocol realizing $\F_{\pws}$ in the presence of malicious adversaries.

\section{System Model}\label{sec:model}

\noindent\textbf{Participants.} %
We work in a setting which involves three semi-honest entities: (1) the users, 
(2) the group manager, and (3) the Web search engine.
More specifically, The \emph{users} are the individuals who submit query
  terms to the search engine and who wish to prevent the search engine
  from building user profiles. We use $u$ to denote a user.
The role of the \emph{group manager}, denoted
  by $G$, is to group users for execution of the protocol introduced above.  
The \emph{Web search engine} provider, denoted by $W$, 
is the entity that provides  a list of best-matching Web pages, 
usually accompanied by a short summary and/or parts of the document. 
 Note that a search engine has no incentive to protect the users' privacy.

\smallskip
\noindent\textbf{Communication channels.} %
For our communication model, we assume that there is a broadcast channel
whereby  users send messages to all other users in a single round.
However, sometimes the network may not have a broadcast channel, 
and we will need to emulate a broadcast channel using point-to-point communication.

\smallskip
\noindent\textbf{Restrictions on adversary.} %
As in the security model, we consider that an adversary is not allowed to break current encryption schemes that are
computationally secure. 
Further, we assume that, for correctness, there is at least one honest user in a group. 
In contrast to~\cite{CVH09}, we allow collusion between 
two different participants in the protocol. 
However, our solution does not consider several desirable properties for efficiency reasons, namely 
abrupt termination, fairness, and guaranteed output delivery.
The first of these properties has been relatively better studied than the remainder.
For example, Garg et al.~\cite{GGHR14} constructed a two-round protocol for general computation.
However, for protocols that consider fairness and guaranteed output delivery, our understanding of round complexity is still incomplete.

\section{Our Shuffle Protocol for Web Queries}\label{sec:pws}

In this section, we describe the concrete construction of our CP-PWS system. 
We begin by describing the model of execution for our PWS construction
 and its participants, explaining the concepts of queries, shuffle, and cryptographic keys.  
 We then give the description of our scheme. 
 Finally, we discuss security and  efficiency issues.

\begin{figure*}[t]
\begin{center}
\fbox{
\begin{minipage}[c]{0.9\textwidth}
\textbf{protocol} Private Web Search

\renewcommand{\labelitemi}{--}
\begin{itemize}
\item Inputs: a list of query terms $(\qry_1,\qry_2,\ldots,\qry_n)$ from each user
\item Auxiliary inputs: a security parameter and a group size $n$
\item The protocol actions:
\begin{enumerate}
\item \textit{Setup}. This step consists of two major tasks as follows:
\renewcommand{\labelitemii}{$\bullet$}
\begin{itemize}
\setlength{\itemindent}{0em}
\item Group setup: A group  of $n$ users is created, 
	and a group information is published. 
\item Parameter selection: $N$ group managers jointly generate  parameters 
	for  threshold cryptographic primitives
	 and publicize the parameters to the users.
\end{itemize}

\item \textit{Mixing query}. In this step, the users perform the following:
	\begin{enumerate}
	\setlength{\itemindent}{0.7em}
	\item Splitting the query term $\qry_i$ into $n$ shares using
          Shamir's secret sharing scheme.
	\item Encrypting the shares into a list of ciphertexts under the public key of group.
	\item Broadcasting the list of ciphertexts.
	\item Re-encrypting and mixing the received list of ciphertexts.
	\item Sending the updated list of ciphertexts to the group managers.
	\end{enumerate}

\item \textit{Submitting query}. 
	In this step, the group managers jointly recover all query terms 
	by applying the Lagrange interpolation  to the  decrypted lists, and
	submit the recovered  terms to the search engine
	on behalf of the group users.
\end{enumerate}
\end{itemize}

\end{minipage}
}
\end{center}
\caption{An algorithmic description of our protocol $\Pi_\pws$}\label{fig-our-protocol}
\end{figure*}

\subsection{Our Basic CB-PWS Protocol}\label{sec:our-pws}

As mentioned above, the key idea behind CB-PWS solutions is 
for a group to submit a set of search words on behalf of individual users. 
Following this design philosophy, we describe how our protocol 
hides the link between users and their query terms.
We begin by giving a high-level overview of our protocol.

\smallskip
\noindent\textbf{Overview}.  
Our PWS scheme, denoted by $\Pi_\pws$, is divided logically into three phases:
\begin{enumerate}
\item \emph{Setup}. The main goal of the Setup phase is to create a group of users 
who would like to make searches via the search engine. 
In addition, all system parameters for Shamir's secret sharing scheme 
and a public-key encryption scheme will be published to all users in the group.

\item \emph{Mixing query}. Upon completing this phase, 
	all users will hold a reencrypted and permuted version of the distributed query terms.

\item \emph{Submitting query}. The group manager receives a set of queries, 
	without knowing who submitted which query. 
	It then submits the queries to the search engine. 
	Upon receiving a set of query results from the search engine, it broadcasts the result.
\end{enumerate}

We provide an abstract description of our proposal in Figure~\ref{fig-our-protocol}, which summarizes all interactions between the entities 
(e.g., users and the group manager). From this, we can estimate approximately the complexity results.

\subsubsection{Setup} \label{subsubsec:setup}

Let $n$ be the size of the group and let $N$ be the number of
group managers. For convenience, we simply assume that
the group managers know  $n$ and that all users know where 
the group managers are and how to contact them. 
Further, all messages are assumed to be encoded automatically into 
the working group of a given encryption scheme.

Let $\pke=(\tkg,\enc,\tdec,\re)$ be a semantically secure threshold GHE scheme, 
and let  $K$ be a public parameter specifying  Shamir's secret sharing scheme. 
The setup phase then comprises three main activities:

\begin{enumerate}
\renewcommand{\labelenumi}{(\arabic{enumi})}
\item The $N$ group managers $G_1,\ldots,G_N$ collaboratively run the key generation $\tkg(1^\spara,N$) and 
	publish   the system parameters, including the public key $\pk=y$, and with 
	 the parameter $K$, within the restrictions specified in \S\ref{subsec:compt}.
	
\item The $N$ group managers elect a group-manager leader, denoted by $G\in\{G_1,\ldots,G_N\}$, and 
	publicize $G$ to users as a representative of the group managers.

\item When the leading group manager $G$ receives $n$ requests for a private
  query, it responds to all $n$ users indicating the group size of
  $n$. The group manager then constructs a group
  $\{u_1,u_2,\ldots,u_n\}$ and publishes the group
  information, including  the   group name, a list of participating users, 
  and each user's label.  

\end{enumerate}

We defer the details of  techniques for choosing the leader of the group managers and for 
  group construction to Section~\ref{sec:group}.
Note that these subprotocols require only a constant number of rounds.

\subsubsection{Mixing Query}\label{subsubsec:mix}

Let $\PT$ be the plaintext domain of $\pke$ and let $F_K$ 
be the working domain for Shamir's sharing scheme such that $F_K\subset \PT.$
Let $\qry_i\in F_K$ be a query term from $u_i$.
After receiving both the group information and 
 the system parameters from $G$ as a response to its query request, 
 each user $u_{i\in[n]}$ performs six steps:

\begin{enumerate}\itemsep2mm
\renewcommand{\labelenumi}{(\arabic{enumi})}

\item Chooses a set of  random coefficients $\{r_{i,k}\}_{k=1}^{n-1}$ such that $r_{i,k}\in F_K$ 
and $|r_{i,k}|=\lceil\log\tilde{q}\rceil$, and determines 
\begin{equation*}
R_i(x)=\sum_{k=1}^{n-1}r_{i,k}x^{k}+\qry_i\in {F_K}^{n-1}[x]
\end{equation*}

\item Computes the shares of the query term
$\mathsf{q}_i$  by evaluating $R_i$ at each point $j\in[n]$
and sets ${v}_{i,j}= R_i(j)\in F_K$ for each $i,j\in [n]$. 

\item Generates a random integer $\alpha_i$ such that $|\alpha_i|=2\lceil\log n\rceil$
and, for all $j\in[n]$, defines the shares by appending it to each of them as
$\mathsf{v}_{i,j} = {v}_{i,j}\parallel \alpha_i.$ 
We then have $|\mathsf{v}_{i,j}|=\lceil\log \tilde{q}\rceil+2\lceil\log n\rceil\leq \lceil\log q\rceil.$

\item Computes $\bar{\mathsf{v}}_{i,j}=\enc_\pk(\mathsf{v}_{i,j})$  for each $j\in[n]$, 
and broadcasts a list  
$\langle i,j,\bar{\mathsf{v}}_{i,j}\rangle_{j\in[n]\backslash\{i\}}$ 
to all other users.

\item Because $u_i$ can build  the array of 4-tuples: 
\begin{equation*}
\left[
\begin{array}{ccccc}
\perp &\cdots  &\langle 1,i,\ev_{1,i}\rangle & \cdots & \langle 1,n,\ev_{1,n}\rangle\\
\langle 2,1,\ev_{2,1}\rangle & \cdots &\langle 2,i,\ev_{2,i}\rangle & \cdots & \langle 2,n,\ev_{2,n}\rangle\\
\multicolumn{5}{c}{$\vdots$}\\
\langle n,1,\ev_{n,1}\rangle & \cdots &\langle n,i,\ev_{n,i}\rangle & \cdots & \perp\\'
\end{array}
\right]
\end{equation*}
where $\perp$ indicates that a 4-tuple with an encrypted share is unknown to the corresponding cell,
it sets $\ct{\boldsymbol{v}}_i=(\ct{\mathsf{v}}_{1,i},\ct{\mathsf{v}}_{2,i},\ldots,\ct{\mathsf{v}}_{n,i})$ and then
computes a rerandomized and shuffled version of $\ct{\boldsymbol{v}}_i$: 
\begin{equation*}
\rct{\boldsymbol{v}}_i=(\rct{\mathsf{v}}_{1,i},\rct{\mathsf{v}}_{2,i},\ldots,\rct{\mathsf{v}}_{n,i}),
\end{equation*} 
where, for each $\ell\in[n]$,  
$\rct{\mathsf{v}}_{\ell,i}=\re_\pk\left(\bar{\mathsf{v}}_{\pi_i(\ell),i};\gamma_\ell\right)$ 
with $\pi_i$  is a random permutation on $[n]$ and $\gamma_{i,\ell}\in\Z_q^\times$.

\item Sends the new list $\rct{\boldsymbol{v}}_i$ to the group manager $G$.
\end{enumerate}

\subsubsection{Submitting Query} \label{subsubsec:submit}

The group manager performs the following four steps:	\hfill

\begin{enumerate}\itemsep1mm
\renewcommand{\labelenumi}{(\arabic{enumi})}
\item Constructs  an $n\times n$ matrix $M$ 
by decrypting all of the received ciphertext vectors:

\begin{equation*}
M  = 
\begin{bmatrix}
\mathsf{v}_{\pi_1(1),1} & \mathsf{v}_{\pi_1(2),1} &  \cdots & \mathsf{v}_{\pi_1(n),1}\\
\mathsf{v}_{\pi_2(1),2} & \mathsf{v}_{\pi_2(2),2} &  \cdots & \mathsf{v}_{\pi_2(n),2}\\
\vdots	       &    \vdots      & \vdots &  \vdots \\
\mathsf{v}_{\pi_n(1),n} & \mathsf{v}_{\pi_n(2),n} &  \cdots & \mathsf{v}_{\pi_n(n),n}\\
\end{bmatrix}
\end{equation*}

\item  Parses each $\mathsf{v}_{\pi_i(j),i}$ into a share of a query term and a random padding,
 and collects only the  $n$ shares as
 \begin{equation*}
\boldsymbol{v}_i= \left\{v_{i,1},v_{i,2},\ldots,v_{i,n}\right\},
 \end{equation*} 
 where the random padding is the same for all shares. 

\item Reconstructs the polynomial $R_i(x)$ of degree $n-1$ 
	using the Lagrange interpolant and $n$ points in $\boldsymbol{v}_i$,
	for all $i\in[n]$, and 
	recovers a list of query terms $Q=\{\qry_1,\qry_2,\ldots,\qry_n\}$
	from the fact that $\qry_i$ is the constant term of $R_i(x)$ for each $i\in[n]$.

\item Submits the set $Q$ to the search engine $W$ and then 
	broadcasts to the users a  result  set $\{a_1,\ldots,a_n\}$ received from $W$. 
\end{enumerate}

In the following subsection, we analyze the performance of our construction in
terms of its efficiency. 
Next, we analyze the security of our protocol by examining its various behaviors.

\subsection{Performance}\label{subsec:performance-anal}

Our protocol $\Pi_\pws$  is compared to other CB-PWS solutions in terms of
three efficiency measures: computation, communication, and rounds. For
this purpose, we first analyze the performance of our proposal. Then,
the schemes proposed by Castell\`{a}-Roca et al.~\cite{CVH09} 
and Lindell and Waisbard~\cite{LW10} are compared to
our proposal. Because the scheme by Romero-Tris et al.~\cite{RCV11} is identical
to Castell\`{a}-Roca et al.'s scheme except for adding ZKPs 
and replacing the shuffling by a permutation
network, we omit Romero-Tris et al.'s scheme from the comparison. 
Furthermore, we are unable to provide a fair comparison between the scheme proposed
by Kim and Kim~\cite{KK12} and our scheme because  their scheme 
restricts the  query term size to $\frac{\log q}{n}$, i.e., $|\qry| \leq\frac{\log q}{n}$.

\subsubsection{Computation \& Communication Complexity} \label{subsubsec:cc-cpx}

We first analyze the computation and communication costs for running the protocol $\Pi_\pws$. 
For a fair comparison, 
we assume that our construction also employs an El Gamal encryption scheme
whose working group is $\G_q$ of order $q$, which is the subgroup of $\Z_p^\times$ for 
primes $q,p$. For example, $p$ may be 1024 bits long while $q$ is 512 bits long.
We denote by $\mexp(\ell)$ a modular exponentiation
 of an $\ell$-bit integer and by $\mmul(\ell)$ a modular multiplication
 of two $\ell$-bit integers. We note that, because Lindell and Waisbard's scheme
 uses a two-layer encryption where Cramer and Shoup's cryptosystem~\cite{CS98}
 takes as input an El Gamal ciphertext, most of their modular
exponentiations   should be carried out modulo a
2048-bit integer rather than a 1024-bit integer.

\smallskip
\noindent\textbf{Computations}. Table~\ref{table:comp-cpx} presents a comparison of computations 
for our scheme and our competitors. For a fair comparison, we did not count 
the number of cryptographic  operations against a malicious adversary, 
e.g., ZKPs.

\begin{table}[h!]
\renewcommand{\arraystretch}{1.25}
\centering
\caption{Comparisons of computation costs}\label{table:comp-cpx}
\begin{tabular}{|@{ }c@{ }|c|c|} \cline{2-3}
 \multicolumn{1}{c|}{} & Modular Exponentiations & Modular Multiplications \\ \hline
 Ours & $4n\mexp(\lceil\log q\rceil)$& $\left(3n+O(n\log^2n)\right)\mmul(\lceil\log q\rceil)$\\ \hline
 \multirow{2}{*}{\cite{LW10}} & $(n+2)\mexp(\lceil\log q\rceil)+$ & $(n+1)\mmul(\lceil\log q\rceil)+$ \\ 
		& $11n\mexp(2\lceil\log q\rceil)$	& $6n\mmul(2\lceil\log q\rceil)$ \\ \hline
 \cite{CVH09} & $(3n+2)\mexp(\lceil\log q\rceil)$ & $(3n+1)\mmul(\lceil\log q\rceil)$\\ \hline
 
\end{tabular}
\end{table}

The `Mixing Query' step in our scheme requires evaluation of a polynomial at many points.
It is well known that a fast evaluation of a polynomial in $F_K[x]$, of degree less than $n$ and for $n$ points in $F_K$,
can be performed using at most $O(n\log^2 n)$ operations in $F_K$.
As a result, the computation complexity of our protocol is in total 
$O(n)$ modular exponentiations in addition to $O(n\log^2n)$  modular multiplications.

\smallskip
\noindent\textbf{Communications}. In our scheme, each user first needs to send $(n-1)$ El Gamal ciphertexts
and then $n$ El Gamal ciphertexts, where each El Gamal ciphertext comprises two $\lceil\log p\rceil$-bit integers. 
Because of the $n$-layered encryption in the Lindell and Waisbard scheme, the ciphertext size  may grow 
exponentially. To avoid this problem, the authors apply a hybrid encryption technique by 
introducing a  secure block cipher  that can process a $\kappa$-bit block in one operation.
In their scheme, the $i$-th user outputs an $i$-layered Cramer-Shoup ciphertext whose size is 
$(i\cdot 4\cdot 2\lceil\log p\rceil+\kappa)$ bits. For $n$ users, the  transmission size 
becomes $\frac{n(n+1)}{2}\cdot 4\cdot 2\lceil\log p\rceil+n\kappa=4n(n+1)\lceil\log p\rceil+n\kappa$ bits.
Because each user holds $n$ of these ciphertexts, the transmission 
requires $4n^2(n+1)\lceil\log p\rceil+n^2\kappa$ bits in total. Furthermore, each user 
sends $n\lceil\log p\rceil$ bits for decryption to other users.

\begin{table}[h!]
\renewcommand{\arraystretch}{1.25}
\centering
\caption{Comparisons of communication costs}\label{table:comm-cpx}
\begin{tabular}{|c|c|} \cline{2-2}
 \multicolumn{1}{c|}{} & Transmissions (in bits) \\ \hline
 Ours & $2n(2n-1)\lceil\log p\rceil$\\ \hline
 \cite{LW10} &  $4n^2(n+1)\lceil\log p\rceil+n^2\kappa+n^2\lceil\log p\rceil$ \\ \hline
 \cite{CVH09} & $3n^2\lceil\log p\rceil$\\ \hline
 
\end{tabular}
\end{table}

In consequence, our scheme has $O(n^2\lceil\log p\rceil)$ communication complexity, 
whereas Lindell and Waisbard's scheme has $O(n^3\lceil\log p\rceil)$ complexity.

\subsubsection{Round Complexity} \label{subsubsec:rnd-cpx}

We have yet to show that the total number of rounds for running 
our PWS solution  is constant
because  we do not know the round number required to 
build a group of $n$ users.

However, if building a group of users can be performed with a constant number of rounds, 
it is clear that our protocol will have a constant round complexity. 
The main part of our protocol comprises  four  rounds: 
(1)  applying Shamir's secret sharing and encryption to a query term, and broadcasting the resulting list, 
(2) shuffling the resulting set and sending it to the group manager,
(3) recovering a set of query terms and submitting them to a search engine, and
(4) broadcasting the search results to the users.  
In contrast, other CB-PWS proposals have $O(n)$-round complexity.

\subsection{Security} \label{subsubsec:sec-hbc}
We continue by arguing that our protocol $\Pi_\pws$ is secure in terms of the semi-honest model.

Assume that there exists at least one honest group manager in $\{G_1,\ldots,G_N\}$.
Because all private inputs are encrypted by semantically secure encryption $\enc_\pk(\cdot)$,
no users can learn nontrivial information about other users' private query values
during the protocol $\Pi_\pws$ with any significant probability. 
Because a random permutation of an honest user  is secret to the adversary and,
in this security model, the adversary should follow the instructions of the protocol,
the protocol leaks no nontrivial information, even for the case of a conspiracy between a group manager and 
a set of $t$ corrupted users, where $t<n$.

If all group managers are corrupt, then the adversary could identify which query term 
has been submitted by which user. This is because each share of a query term carries a random padding,
and therefore decrypting an El Gamal ciphertext originating from a specific user would
compromise the unlinkability of our protocol.

\subsection{Upgrading to Malicious Security}{\label{subsec:sec-anal}

A standard technique for ensuring the security of a cryptographic protocol is 
to show the achievability of simulation-based security of the protocol.
To achieve our security goal against malicious entities, 
we will need to utilize cryptographic tools as subprotocols.  
The first tool is a ZKP of knowledge about a discrete logarithm (DL).
There have been many efforts to construct 
a ZKP for the language of the nonzero exponent of $g$:
\begin{equation*}
\lang_\podl=\{(\pk,h)|\exists x \text{ s.t. } h=g^x\}
\end{equation*}
We use a ZKP protocol designed for a signature scheme by Schnorr~\cite{Sch89},
denoted by $\podl_\pk\{(x)|g^x\}.$

The second tool is  a ZKP that, given a public ciphertext, a
user knows the corresponding plaintext. 
We use standard techniques to design a ZKP protocol for the language:
\begin{equation*}
\lang_\popk=\left\{(\pk,\ct{v})|\exists\ v\in\G_q\text{ s.t. }\ct{v}=\enc_\pk(v)\right\},
\end{equation*} 
and we denote by $\popk_\pk\{(v)|\ct{v}=\enc_\pk(v)\}$ a ZKP from this subroutine. 
ZKP techniques for proving plaintext knowledge have been well studied elsewhere~\cite{CDN01}.

The final tool is a verifiable shuffle that enables proof of 
the correctness of a shuffle, where a shuffle of ciphertexts $\ct{\bs{v}}=(\ct{v}_1,\ldots,\ct{v}_n)$
 is a new set of ciphertexts $\rct{\bs{v}}=(\rct{v}_1,\ldots, \rct{v}_n)$ with the same plaintexts in permuted order.
We also use a ZKP protocol for the language of the shuffle of $2n$ El Gamal encryptions:
\begin{equation*}
\lang_\vs=\{(pk,\ct{\bs{v}},\rct{\bs{v}})|\exists\ (\pi,\bs{\gamma})\text{ s.t. }\forall\ i\in[n],\rct{v}_{\pi(i)}=\re_\pk(\ct{v}_i;\gamma_i)\},
\end{equation*}
where $\pi$ is a random permutation on $[n]$ and 
a random vector $\bs{\gamma}=(\gamma_1,\ldots,\gamma_n)\xleftarrow{\text{\$}}(\Z_q^\times)^n$. 
$\vs_\pk\{(\pi,\bs{\gamma})|\rct{v}_{\pi(i)}=\re_\pk(\ct{v}_i;\gamma_i)
\wedge \ct{v}_i\in\ct{\bs{v}}\wedge \rct{v}_{\pi(i)}\in\rct{\bs{v}}\}$ 
is shorthand notation for a ZKP that a user knows a witness
$(\pi,\bs{\gamma})$ to the correctness of a shuffle. 
There is a considerable literature on this topic. For example, 
VS protocols by Groth~\cite{Gro10} and Neff~\cite{Neff01} are well known. 
Note that all values not enclosed by () are assumed to be known to the verifier.

\subsubsection{Our CB-PWS Protocol for the Malicious case}\label{subsubsec:fixing}

Using  the above subprotocols,  we can strengthen the basic protocol described in \S\ref{sec:our-pws} 
so that the modified protocol is secure against a malicious adversary. 
We denote this extended version as $\Pi^\star_\pws$.
The remainder of this section deals with steps that should be modified for stronger security.
To avoid repeating the basic descriptions, we will describe just the required modifications to the three phases:

\smallskip
\noindent\textbf{Setup}.  %
We need to change only Step (1). Specifically,
each group manager $G_{i\in [N]}$ chooses a random private key $x_i\xleftarrow{\text{\$}}\Z_q^\times$
and publishes its public-key share $y_i=g^{x_i}$ along with a ZKP of 
knowledge of $y_i$'s discrete logarithm using $\podl_\pk\{(x_i)|g^{x_i}\}$.

\smallskip
\noindent\textbf{Mixing Query}. %
We make no changes in Steps (1) to (3) of this phase.

In Step (4), each user $u_i$ computes $\ct{\mathsf{v}}_{i,j}$  
 with proof of plaintext knowledge: 
\begin{equation*} 
 \popk_\pk\{(\mathsf{v}_{i,j})|
 \ct{\mathsf{v}}_{i,j}=\enc_\pk(\mathsf{v}_{i,j})\},
\end{equation*}
for each $j\in[n]$ and sends it to all other users.

In Step (5), on receiving an encrypted share, each user 
verifies proofs of plaintext knowledge. 
If this check fails, the protocol is terminated.
If all verifications of proofs from other users are valid, 
for a ciphertext vector $\ct{\bs{v}}_i=(\ct{\mathsf{v}}_{1,i},\ldots,\ct{\mathsf{v}}_{n,i})$,
each user computes a rerandomized and shuffled version of this vector,
$\rct{\bs{v}}_i=(\rct{\mathsf{v}}_{1,i},\ldots,\rct{\mathsf{v}}_{n,i})$
together with proofs of correct shuffle, for each $\ell\in[n]$:
\begin{equation*}
\vs_\pk\{(\pi_i,\bs{\gamma}_{i})|\rct{\mathsf{v}}_{\pi_i(\ell),i}=\re_\pk(\ct{\mathsf{v}}_{\ell,i};\gamma_{i,\ell})\},
\end{equation*} 
using  a vector of randomizers $\bs{\gamma}_i=(\gamma_{i,1},\ldots,\gamma_{i,n})$ 
and a random permutation $\pi_i$ of its own choice,
where $\ct{\mathsf{v}}_{\ell,i}$ is a component of $\ct{\bs{v}}_i$ and 
$\rct{\mathsf{v}}_{\pi_i(\ell),i}$ is a component of $\rct{\bs{v}}_i$.
The user then sends it to the group manager $G$ in Step (6).

\smallskip
\noindent\textbf{Submitting Query}. %
On receiving a vector of ciphertexts $\rct{\bs{v}}_i$ from a user $u_i$, 
the group manager verifies the proofs of correct shuffle.
If this verification fails, it terminates the protocol.
Only after verification from all users, it constructs the matrix $M$ in Step (1).

In Step (2), the group manager receives a vector of $n$ shares $\bs{v}_i=(v_{i,1},\ldots,v_{i,n})$
such that all $v_{i,j}$'s have the same random padding $\alpha_i$,
without knowing which user submitted the vector $\bs{v}_i$.

The remaining two steps are unchanged.

\subsubsection{Security Proof for the Malicious Case}\label{subsubsec:mal-sec}

Before considering the proof of security,
we observe that if all users and a group manager are honest, 
$\usr_i$ outputs a set of query results $\{a_1,\ldots,a_n\}$
with a probability greater than $\frac{1}{n^4}$.
In this case, if  a random padding $\alpha_i\in\{0,1\}^{n^2}$ is unique in an execution of our protocol, then, 
because of the Lagrange interpolation, a unique polynomial is reconstructed
from the shares with the same $\alpha_i$ and a query term $\qry_i$ is correctly recovered.
However, for the case of $\alpha_i=\alpha_{j\neq i}$, the group manager 
drops both query terms  $\qry_i,\qry_j$. 
Therefore, the query term $\qry_i$ cannot be recovered with  probability $\Pr[\alpha_i=\alpha_j]\leq \frac{1}{n^4}$.
For example, for a modest group size of $n=30$, the probability is $\frac{1}{30^4}\approx2^{-20}$.

We now prove that our protocol $\Pi^\star_\pws$ satisfies the security definition
given in Definition~\ref{df:sec}.

\begin{thm}\label{thm:security}
Assume that at least one group manager is honest, 
and  that $\podl,\popk,$ and $\vs$ are the ZKPs for $\lang_\podl,\lang_\popk,$ and 
$\lang_\vs$, respectively, and $(\tkg,\enc,\tdec,\re)$ is the threshold El Gamal encryption scheme with 
semantic security.
Then our protocol $\Pi^\star_\pws$ for any coalition $J$ of colluding users such that $|J|<n$
 securely computes $\mathcal{F}_\pws$
in the presence of malicious adversaries, assuming a broadcast channel.
\end{thm}

\begin{IEEEproof}
We prove the security by constructing an algorithm $\Sim$, called a simulator, for an adversary in the ideal model.
The simulator works within the ideal model, but it interacts with the corrupted users in $J$ 
without $\A$  detecting that it is not within the real model. The trusted party takes the input from $\Sim$ and the honest users, 
and gives the query result set $\Sim$ to the honest users.
The simulator $\Sim$ then interacts with the malicious users in $J$, pretending to be one or more honest users,
enabling them to know the query result set.
For the trivial case of $J=\varnothing$, 
we have demonstrated that the output is correct.
Our proof is in a hybrid model where   a trusted party runs the ZKPs
of knowledge for $\lang_\podl,\lang_{\popk},$ and $\lang_\vs$.

The simulator $\Sim$ proceeds as follows:
\begin{enumerate}
\item $\Sim$ performs the \emph{Setup} phase  as follows:
	\begin{enumerate}
	\item For each honest group manager $G$, $\Sim$ chooses a uniformly random $x_G\in\Z_q^\times$.
		$\Sim$ sends $y_G=g^{x_G}$ to $\A$ and emulates the ideal functionality of $\lang_\podl$ 
		by sending $x_G$ to $\A$. 
	\item For each malicious group manager $\tilde{G}$, $\Sim$ receives from $\A$,
		$(y_{\tilde{G}},x_{\tilde{G}})$ for the ideal functionality of $\lang_\podl$ and 
		records $x_{\tilde{G}}$ only when $y_{\tilde{G}}=g^{x_{\tilde{G}}}$.
		Otherwise, it aborts, sending an error message to the trusted party for $\F_\pws$.
	\end{enumerate}
\item Let $I=[n]\backslash J$. For each simulated honest user $\usr_{i\in I}$, the simulator $\Sim$:
	\begin{enumerate}
	\item chooses a random polynomial $R_i(x)$ such that all coefficients are 
		in $F_{K}$ and its constant term is a query word $\qry_i$ of its choice.
	\item chooses random paddings $\alpha_i$  and 
		constructs encryptions of shares with them.
	\end{enumerate}
\item To perform the \emph{Mixing Query} phase of the protocol, the simulator $\Sim$:
	\begin{enumerate}
	\item sends the encrypted shares to all malicious users in $J$
		and simulates the ideal functionality of $\lang_\popk$ by sending each share.
	\item receives from $\A$ a list of encrypted shares for each malicious user $\usr_{\xi\in J}$.
	\item receives  from $\A$ its input $V_\xi=(\mathsf{v}_{\xi,j})_{j\in[n]\backslash\{\xi\}}$ 
		 for the ideal implementation $\lang_\popk$
		for each malicious user $\usr_{\xi\in J}$. If $\A$ fails to prove its correctness, then $\Sim$ 
		sends an error message to the trusted party $\F_\pws$ and aborts.
	\item emulates the ideal functionality of $\lang_\vs$ as follows:
		it constructs an $n\times n$ matrix $M_i$ for each honest user $\usr_i$ in $I$
		and sets its $i$-th column vector to $\ct{\bs{v}}_i$.
	 	For all $i\in I$, $\Sim$ chooses a random vector $\bs{\gamma}_i=(\gamma_{i,1},\ldots,\gamma_{i,n})\xleftarrow{\text{\$}}(\Z_q^\times)^n$ 
		and  a random permutation $\pi_i$ over $[n]$ and 
		computes a vector $\rct{\bs{v}}_i$. $\Sim$ sends $\rct{\bs{v}}_i$ to $\A$ and emulates the ideal functionality of $\lang_\vs$ 
		by sending the vector $\bs{\gamma}_{i}$ and the permutation $\pi_i$.
	\item receives from $\A$ a vector of mixed encryptions $\rct{\bs{v}}_\xi$ for each malicious user $\usr_{\xi\in  J}$.
	\item receives from $\A$ its input $(\pi_\xi,\bs{\gamma}_\xi)$ for the ideal functionality of $\lang_\vs$
		for each malicious user $\usr_\xi$. $\Sim$ communicates with $\A$ to check 
		whether its proof is valid or not; if not, then $\Sim$ sends an error message to $\F_\pws$ and aborts.
	\end{enumerate}
\item To perform the  \emph{Submitting Query} phase, $\Sim$ emulates a group manager $G$: 
	\begin{enumerate}
	\item $\Sim$ sends each vector $\rct{\bs{v}}_{i\in I}$ for all honest users 
	and $\rct{\bs{v}}_{\xi\in J}$ from $\A$ to the group manager.
	\item For each $\xi\in J$, $\Sim$ extracts  $\{\mathsf{v}_{j,\xi}\}_{j\in[n]}$ from 
		the encryption vector $\bs{\rct{v}}_\xi$ using $(\cdot,\bs{\gamma}_\xi)$  in Step 2-f.
	\item Using $V_\xi$ in Step 2-c and $(\pi_\xi,\cdot)$ in Step 2-f, $\Sim$ 
		constructs a vector $\bs{v}_\xi=(\mathsf{v}_{\xi,j})_{j\in[n]}$ and
	 	calculates $\qry_\xi$ from the vector using Lagrange interpolation. 
	\item $\Sim$ sends a set of query terms $Q=\{\qry_\xi\}_{\xi\in J}\cup\{\qry_i\}_{i\in I}$
	to the trusted party $\F_\pws$ and receives as the answer a set $\{a_1,\ldots,a_n\}$.
	\end{enumerate}
\item $\Sim$ outputs whatever $\A$ does.
\end{enumerate}

We can see that the simulator runs in polynomial time, and 
the malicious users cannot identify that they are communicating with the simulator, 
which is working with the ideal model, rather than with other users in the real model.
The correct answer is learned by all users for both real and ideal models.
Therefore, we may conclude the theorem.
\end{IEEEproof}

\smallskip
\noindent\textbf{Limitations.} %
To maintain security against malicious adversaries,
we require that all participants prove the correctness of each protocol step.
The above subprotocols can be proven correct by using only 
so-called $\Sigma$-protocols, which need just three rounds of interaction~\cite{Dam02,CDS94}.
$\Sigma$-protocols are not known to be zero-knowledge, but they 
enable efficient proofs of correctness for the protocol steps.
Applying the Fiat-Shamir heuristic~\cite{FS86} enables 
the obtained proof to be zero-knowledge with a single message, but for the \emph{random oracle} model.

If a broadcast channel is not available, we need to modify Theorem~\ref{thm:security}.
First, the number of corrupted users $t$ should be strictly bound.
Because we are working in an authenticated setting, we are allowed to set $t<\frac{n}{2}$.
We can then emulate the broadcast channel with \emph{expected constant rounds} using 
point-to-point interaction.~\cite{KK06}
The round overhead of our protocol in this setting will be dominated by this  cost.

\section{Group Construction}\label{sec:group}

We have a variety of approaches to ensuring the establishment of an anonymous channel between users and search engines (e.g., the PIR technique~\cite{CGKS95} and Chaum's mix-net~\cite{Cha81}). In principle, they can be applied to obtain a secure PWS solution. 
The most convincing reason for their applicability to PWS solutions is that they require users to engage in a group, enabling the users to hide their identity within the group. The larger the group, the greater is the anonymity available to users. Despite the importance of the group setup issue, many previous CB-PWS approaches assume a particular group or do not offer a specific group-setup protocol.

Unlike other approaches in the CB-PWS literature, we pay attention to developing a set of specific protocol activities for users to create a group of a fixed size and to elect a group manager for the group. In the remainder of this section, we present our group-setup protocol and analyze its security for the random oracle model. To start, we summarize the relevant abbreviations and notations in Table~\ref{table:recap}.

\begin{table}[!h]
\renewcommand{\arraystretch}{1.2}
\centering
\caption{Notation and abbreviations}\label{table:recap}
\tabulinesep=0.4mm
\begin{tabu}{r|l}\hline
$n_u$ & the total number of users in the system\\ 
$n$ & the group size \\
$n_g$ & the number of groups, i.e., $n_g=\lfloor\frac{n_u}{n}\rfloor$\\
$t$ & the number of corrupted users \\
$\{0,1\}^*$ & the set of all binary strings \\
$\{0,1\}^\alpha$ & the set of all binary strings of length $\alpha$\\
$s_1\parallel s_2$ & the concatenation of strings $s_1,s_2\in\{0,1\}^*$\\
$J$ & an index set of corrupted users in a group\\
$I$ & an index set of honest users in a group\\
$\gp_j$ & the $j^\text{th}$ group of size $n$  in the system\\ \hdashline[1pt/1pt]
\multicolumn{2}{l}{%
$\left.\begin{aligned}%
H_1:\{0,1\}^*\rightarrow\{0,1\}^M \\
H_2:\{0,1\}^{M\cdot(n_u)}\rightarrow\{0,1\}^{n_u\cdot\log (n_u)}\\
H_3:\{0,1\}^{\log (n_u)+n_u|r|}\rightarrow\{0,1\}^{\log (n_g)}
\end{aligned}%
\right\}\text{random oracles}
$
}  \\ \hline
\multicolumn{2}{l}{\scriptsize $M$: a positive integer of polynomial size for $\spara$}\\
\multicolumn{2}{l}{\scriptsize $r$:  a random string in $\G_q$}
\end{tabu}
\end{table}

\subsection{Group Setup}\label{subsec:group-setup}

As for all other protocols in the literature,
we assume the existence of an ideal public bulletin-board (PBB) functionality. In fact, this assumption is equivalent to the assumption of a broadcast channel. 
Informally the information is intentionally published and
recorded permanently on the PBB (e.g., Web pages that serve as a PBB).
Figure~\ref{fig:gp-setup} shows our resulting new protocol for constructing a group of $n$ users.

\begin{figure}[!h]
\centering
\begin{tabular}{llll}\hline
\multicolumn{4}{l}{\textbf{protocol} Group setup} \\ \hline
\multicolumn{4}{l}{\textsc{Goal}: $n$ users build a new group $\gp_j$ in a 2-pass protocol}\\
\multicolumn{4}{l}{\textsc{Result}: a group $\gp_j$ of honest  majority, i.e., $|J|<n/2$}\\
1. & \multicolumn{3}{l}{\textit{Protocol messages}.}\\
    & & \multicolumn{2}{l}{$u_i\rightarrow \text{PBB}$: $\left\langle s_i,x_i=H_1(\text{IP}_i,\text{id}_i,r_i)\right\rangle$}\\
    & & \multicolumn{2}{l}{$u_i\leftarrow \text{PBB}$: $\left\langle j=H_3(y^{(i)},s_1,\ldots,s_{n_u})\right\rangle$}\\ 
2. & \multicolumn{3}{l}{\textit{Protocol actions}. The following steps are performed }\\
    & \multicolumn{3}{l}{each time a user $u_i$ requests. Assume that $n_u$ users}\\
    & \multicolumn{3}{l}{have been registered in the system.}\\
    & (a) & \multicolumn{2}{l}{$u_i$ computes $x_i=H_1(\text{IP}_i,\text{id}_i,r_i)$ where $r_i,s_i$}\\
    &  & \multicolumn{2}{l}{are randomizers from $\G_q$, $\text{IP}_i$ is the IP address,}\\
    &  & \multicolumn{2}{l}{and $\text{id}_i$ is its  ID registered in the system.}\\
    & (b) & \multicolumn{2}{l}{$u_i$ sends $(s_i,x_i)$ to PBB.}\\
    & (c) & \multicolumn{2}{l}{PBB computes $y=H_2(x_1,\ldots,x_{n_u})$ and defines }\\
    &  & \multicolumn{2}{l}{$y:=y^{(1)}\parallel\cdots\parallel y^{(n_u)}$ where $\forall i,|y^{(i)}|=\log n_u.$}\\
    & (d) & \multicolumn{2}{l}{PBB computes $j=H_3(y^{(i)},s_1,\ldots,s_{n_u})$ and }\\
    & & \multicolumn{2}{l}{assigns to $\gp_j$ the user $u_i$.}\\
    & (e) & \multicolumn{2}{l}{PBB sends the $j$ to $u_i$.}\\ \hline
\end{tabular}
\caption{Group setup protocol}\label{fig:gp-setup}
\end{figure}

Our protocol is a variant of Lindell and Waisbard's group-setup protocol in~\cite[\S5.2]{LW10}. 
For several technical reasons,  we need to develop a modified version rather than use  their original protocol.

The primary reason for our modification is that our PWS protocol frequently uses a \emph{broadcast channel} to control the round complexity.  
Because our group-setup protocol should be run before our main CB-PWS protocol, the group-setup protocol might appear to be run in isolation. 
However, we need to emulate the broadcast channel in a point-to-point network using a broadcast protocol. If we wish to realize the broadcast protocol within a fixed number of rounds, we have to strictly restrict  the number of corrupted users $t$.

Classical results (e.g., ~\cite{PSL80,FL82}) show that  achieving broadcast among $n$ users incurs $\Omega(t)$-round complexity only if the number of corrupted users $t$ satisfies $t<n/3$. Fortunately, more recent results based  on randomization techniques show that it is achievable in \emph{expected} constant round time for $t<n/2$, i.e., in an honest-majority setting~\cite{FM97,FG03,KK07}. 
Unlike Lindell and Waisbard's scheme,  our protocol that relies on broadcasting does not allow the adversary to have an index set $J$ such that  $|J|\geq n/2$.
In an authenticated setting like our protocol, 
restricting to a \emph{computationally bounded} adversary  allows a limit of $t<n$.
Therefore, because of the different assumptions in the communication model, we need to devise  a new group-setup protocol.

A second reason relates to practical matters. 
To justify our assumption that $|J|<n/2$ for any index set of corrupted users $J$, we need to guarantee  the probability that the adversary can create a set $J$ whose cardinality $|J|\geq n/2$ is very small, i.e., negligible in $\spara$.  However, our new protocol  should not only consider the strong assumption described above, but also ensure that  the impact of our modification on the probability of maliciously grouping should be minimal, even under  this  assumption. For this reason,  a new random oracle $H_3$ is added to our protocol. 
Because the output of  a random oracle  is uniformly distributed, the additional random oracle $H_3$ makes it harder for the adversary to add members to a group, as it would like.

We now compute the probability that an adversary can maliciously generate a group $\gp_j$ so that  $|\gp_j|$=$n$, but half of its member users  are corrupted (i.e., $n/2$ users are malicious), for some $j\in[n_g]$, where  $n_g=\lfloor\frac{n_u}{n}\rfloor$. We mean by  a  ``malicious'' group that the group contains at least $\frac{n}{2}$ malicious users, with the number of honest users being less than $\frac{n}{2}$. This is clearly an undesirable case because ambiguity with respect to users' query terms will  drastically increase,  particularly when the size of the group is not large.

The following theorem shows that the probability   still remains very small with respect to the security parameter $\spara$. 
We write the probability as $\Pr[\bad]$ in the theorem.

\begin{lem}\label{thm:mal-gp}
Let $n_u$ be the total number of registered users, and let $t$ be the number of corrupted users. Let $n$ be the number of users in a group.
For every positive integer $n_u,n,t\in\N$, and for all PPT adversaries $\A$,  the probability of a bad event is given by: 
\begin{equation*}
\Pr[\bad]\leq \frac{2^n}{n_g}\cdot \left(\frac{n_u-t}{t}\right)^{n/2}\cdot \left(\frac{t}{n_u}\right)^n,
\end{equation*}
where the number of groups $n_g=\lfloor \frac{n_u}{n}\rfloor$.
\end{lem}

\begin{IEEEproof}
Let $\bad$ denote the event that a group is constructed by the adversary in such a way that the group contains $\frac{n}{2}$ corrupted users. 
We then have: 
\begin{IEEEeqnarray*}{rCl}
\Pr[\bad]&=&\frac{1}{\binom{n_g}{1}}\cdot\frac{\binom{t}{n/2}\binom{n_u-t}{n/2}}{\binom{n_u}{n}}=\frac{1}{n_g}\cdot\frac{\binom{t}{\tilde{n}}\binom{n_u-t}{\tilde{n}}}{\binom{n_u}{n}}\nonumber \\
&=& \frac{1}{n_g}\cdot\frac{t!}{(t-\tilde{n})!\tilde{n}!}\cdot\frac{(n_u-t)!}{(n_u-t-\tilde{n})!\tilde{n}!}\cdot\frac{(n_u-n)!n!}{n_u!}\nonumber\\ 
&=&\frac{1}{n_g}\cdot\binom{n}{\tilde{n}}\cdot\frac{\displaystyle\prod_{i=0}^{\tilde{n}-1}(t-i)\prod_{i=0}^{\tilde{n}-1}(n_u-t-i)}{\displaystyle\prod_{i=0}^{n-1}(n_u-i)}\nonumber\\ 
&=&\frac{1}{n_g}\cdot\binom{n}{\tilde{n}}\cdot\frac{\displaystyle\prod_{i=0}^{\tilde{n}-1}(n_u-t-i)}{\displaystyle\prod_{i=\tilde{n}}^{n-1}(t-i)}\cdot\prod_{i=0}^{n-1}\left(\frac{t-i}{n_u-i}\right)\nonumber\\ 
&\leq& \frac{2^n}{n_g}\left(\frac{n_u-t}{t}\right)^{\tilde{n}}\left(\frac{t}{n_u}\right)^n 
\end{IEEEeqnarray*}
The last inequality is derived from  $\binom{k}{\varepsilon k}\leq 2^{k\mathsf{H}(\varepsilon)}$, where $\mathsf{H}(\varepsilon)=-\varepsilon\lg\varepsilon-(1-\varepsilon)\lg(1-\varepsilon)$ for $0\leq\varepsilon\leq 1$, and we assume that $0\lg 0=0$, for convenience.
This completes the proof of the theorem.
\end{IEEEproof}

For example, consider the case that $n_u=10^6,t=10^3,$ and  $n=30$. For this case, we have the probability of a malicious group being smaller than $\frac{30\cdot 10^9}{10^6}\cdot\left(\frac{10^6-10^3}{10^3}\right)^{15}\cdot\left(\frac{10^3}{10^6}\right)^{30}$, which approximates to $10^{-40}.$ It could be argued that this probability is insufficient to protect the users' privacy. However, because the adversary would need to create a malicious group within seconds, we guess that the probability would be meaningful in practice.

\subsection{Group Manager}\label{subsec:gp-manager}

The architectural view of our PWS protocol, described conceptually in Section~\ref{subsec:overview}, is depicted as in Figure~\ref{fig:gp-view}.
The figure  can be considered as a snapshot following construction of a group $\gp_j$ of size $n$ using the group-setup protocol specified in the previous section.

Because the problem we handled in the previous section was the matter of an algorithm to create a group of proper size, 
it is sufficient to find an algorithm and to analyze it.
However, there remain two crucial issues to be addressed.
One issue is the maintenance of a set of group managers, and 
the second is the election of a representative among them.

Concerning maintenance, our idea is to employ the functionality of a PBB.
In a cryptographic sense, a PBB is equivalent to a broadcast channel, 
which has already been discussed in Section~\ref{subsec:group-setup}.
Specifically, the PBB   is used by users to announce their
messages. That is, a message can be posted by any user and read by any other one. 
By saying that a message is ``published'', we mean  that the message appears on the PBB.
Moreover, the published message can not be deleted or modified once posted.
Next, a natural scenario would be that perhaps a dozen users would undertake  voluntarily the role of group manager. 
They could construct a pool of group managers using the group construction protocol 
shown in Figure~\ref{fig:gp-setup}.

%
%

\begin{figure}[!h]
\centering
\begin{tikzpicture}[node distance=0cm,outer sep = 0pt]
\tikzstyle{txt}=[draw=none]
\tikzstyle{rnd}=[draw,circle, outer sep=0pt,inner sep=0pt,minimum height=6mm,minimum width=6mm]
\tikzstyle{msg}=[draw,rectangle,minimum height=0.6cm,minimum width=1.2cm]

  \pgfpathellipse{\pgfpoint{0cm}{0cm}}
                 {\pgfpoint{0.7cm}{0cm}}
                 {\pgfpoint{0cm}{1.3cm}}
  \pgfusepath{draw}
  
  \node[txt] (u1) at (0,1) {\small $u_1$};
   \node[txt] (u2) at (0,0.5) {\small $u_2$};
    \node[txt] (dots) at (0,-0.1) {\footnotesize $\vdots$};
    \node[txt] (un) at (0,-1) {\small $u_n$};
    \node[txt]  at (0,1.6) {\footnotesize group $\gp_j$};
    \node[rnd] (mgr1) at (2.8,1.0) {\footnotesize $G_{i-1}$};
    \node[rnd,thick] (mgr2) at (2.0,0) {\footnotesize $G_i$};
    \node[rnd] (mgr3) at (2.8,-1.0) {\footnotesize $G_{i+1}$};
    \node[rnd] (mgr4) at (4,0.7) {\footnotesize $G_1$};
    \node[rnd] (mgr5) at (4.0,-0.7) {\footnotesize $G_N$};
    \node[txt]  at (5.5,0.13) {\footnotesize search};
    \node[txt]  at (5.5,-0.13) {\footnotesize engine};
    \node[msg] (se) at (5.5, 0) {};
    \draw (mgr1.west) -- (mgr2.north);
    \draw (mgr3.west) -- (mgr2.south);
    \draw[densely dotted] (mgr3.east) -- (mgr5.west);
    \draw[densely dotted] (mgr1.east) -- (mgr4.west);
    \draw (mgr4.south) -- (mgr5.north);
    \draw[<->,thick] (u1) -- (mgr2);
    \draw[<->,thick] (u2) -- (mgr2);
    \draw[<->,thick] (un) -- (mgr2);
    \draw[<->,thick] (se) -- (mgr2);
 \end{tikzpicture}
\caption{Overall communication architecture with grouping}\label{fig:gp-view}
\end{figure}
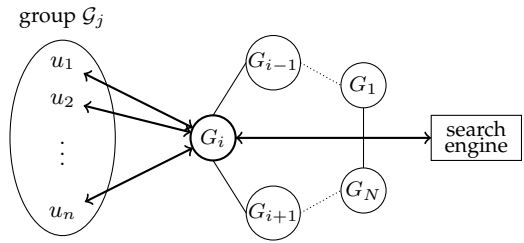

The second issue is the election of a  group manager $G_i$ from the pool $\{G_1,\ldots,G_N\}$.
Leader election plays an important role in the design of fault-tolerant applications. 
There is an extensive literature about this issue (e.g.,~\cite{Sin97,MWV00,ST08}) and 
refer to reference~\cite{Iva13} for a detailed survey of recent results.
 
Intuitively,  a leader can operate as a central coordinator who enforces consistent behavior among users. 
However, in our setting,  the leader of group managers  does not matter if there exists at least one honest group manager.
In other words, a malicious group manager would be allowed to be the leader.
Therefore, we can opt for an auto-election mechanism to choose the group leader after the pool containing honest group managers has been constructed.

For the sake of security, we can use the cryptographic leader-election protocol designed by Katz and Koo~\cite{KK06}.
Their scheme is built on  a moderated VSS,
which adds an extra entity, called a moderator,
to a standard VSS. 
 Their moderated VSS scheme in an authenticated setting (i.e., assuming a public-key infrastructure)
requires only a constant number of rounds, tolerating $t_G<N/2$ malicious group managers~\cite[Corollary 2]{KK06}.
Specifically, the moderated VSS protocol requires thirteen rounds of interaction.
In addition, they provide a two-phase constant-round protocol for leader election in the same setting,
where the round complexity of each phase is determined 
by the underlying moderated VSS~\cite[Corollary 4]{KK06}.

Much simpler leader election protocols could be employed for efficiency reasons.
For example, assuming a trusted leader, we could adapt an efficient variant of Toueg's  constant-round protocol~\cite{Tou84}.

\subsection{Other Considerations}\label{subsec:consider}
We now identify issues that may arise in deploying our protocol 
in a real scenario, discuss the privacy that our protocol covers, and address some other issues. 

\smallskip
\noindent\textbf{Group size.} %
As discussed in Lemma~\ref{thm:mal-gp}, 
our main protocol provides privacy that depends on the group size $n$.
That is, the larger the group, the stronger the privacy that users can obtain. 
However, considering the quality of the Web search service, 
we need to limit the size of groups according to the computing power of users' machines and the acceptable level of latency.
Our example implementation, which used the El Gamal encryption scheme over $1024$-bit $p$ and $512$-bit $q$,
showed that, for $34\leq n\leq 36$, a group manager required approximately 1.02 seconds to submit a list of queries.
We conducted this experiment on an iMac with a
3.4 GHz Core i5 CPU and 16 GB memory.

\smallskip
\noindent\textbf{Static and Dynamic Groups.}  %
The discussion of this matter depends on whether a user or a group manager joins a group.
For users, it might appear that frequently changing groups would achieve higher privacy.
However,  if a unique user identity such as the IP address is repeatedly used in different groups,
the adversary might detect this and have a better chance of designating the user.
Therefore, changing groups, in itself, could lead to loss of privacy.

The primary reason for maintaining multiple group managers is to distribute the key pair in 
the El Gamal encryption scheme. No subsets of group managers whose size is less than a fixed threshold
can then learn user plaintexts.
Both cases seem to imply that any solution involving dynamic groups might be unnecessary in our setting.

\section{Literature Review} \label{sec:related-work}

Balsa et al. divide existing PWS protocols into three classes  according to the underlying technique used to obtain anonymous channels~\cite{BTD12}.  That is, if a PWS solution employs a proxy server to submit  their query words on behalf of users, it is  considered a proxy-based technique. However, the server machine can easily become a target of attackers and has single-point-of-the-failure limitations. A second class of PWS solutions enables users to submit a set of queries so that the real query is well buried within the set. Balsa et al. calls this obfuscation-based PWS. Its technical core is the method that embeds a real query into the set and the indistinguishability level achieved by the method.
The third class, to which our PWS solution belongs, eliminates the possibility of linkability between users and their query terms by using cryptographic tools. We call these schemes  cryptography-based PWS (CB-PWS), following Balsa et al.'s naming convention.


\smallskip
\noindent\textbf{Proxy-based PWS.} %
The first approach is through the use of an anonymous
proxy~(e.g.,~\cite{Ano,Scr,SJBF07}). Users can expect that anonymizers
will prohibit the creation of user profiles through query
unlinkability. There are several options in this category, from simple
mechanisms achieving a low level of anonymity in Web searches
to more reliable but more complicated systems based on onion routing,~\cite{RSG98} such as the Tor network.~\cite{DMS04}
However, the effectiveness of simple solutions is clearly limited. In
addition, as highlighted by~\cite{CVH09}, Tor is not always easy to install and
configure.  Furthermore, it is well known that the HTTP requests
over Tor can become very slow.~\cite{SJBF07} For example, it takes 10 seconds on average to submit a query to Google even when using paths of length 2 (the default length is 3).

\smallskip
\noindent\textbf{Obfuscation-based PWS.} %
Another approach to providing privacy during Web search is based on a
query obfuscation technique
(e.g.,~\cite{ESM06,DSC09,RF10,Tra}). The class of
solutions using query obfuscation involves blending the real queries into a stream of fake queries so that Web search engines cannot create an accurate profile. 
From a privacy point of view, these obfuscation-based solutions have a
critical drawback, namely, that automated queries have  features that are different
from real queries entered by a user, such as randomness. The
authors in~\cite{PS10} demonstrated a classifier implementation that can
distinguish real queries from fake queries generated by
TrackMeNot,~\cite{Tra}  with a mean misclassification rate of only approximately 0.02\%.

\smallskip
\noindent\textbf{Cryptography-based PWS.} %
The third class of solutions involves using  cryptographic algorithms
such as public-key encryption and shuffle.
One of the main advantages of CB-PWS systems over others is that they provide strong privacy guarantees. 
In addition, they are not affected by  the misclassification issue and
are generally faster than anonymizer-based solutions. To our
knowledge, the known solutions can be found in~\cite{CVH09,LW10,RCV11,KK12}. Without loss of generality, because we may consider Romero-Tris et al.'s scheme as a malicious variant of Castell\'{a}-Roca et al.'s scheme,  the differences between the two schemes do not affect the round complexity.

These known solutions utilize the basic idea that, after joining a small group, each  user
encrypts the search query and sends it to other group members. Then,
according to a predefined order, each user provides a shuffled list
of encrypted queries  to his neighbor. The last user broadcasts the final  shuffled version. After group decryption, each user obtains a set of  queries, but cannot know who submitted which query. As a result, Web search engines cannot build accurate user profiles.

We note that the only approach that comes close
to achieving our requirements in the restricted setting is the work by
Kim et al.~\cite{KK12}. The authors proposed a round-efficient CB-PWS scheme based on the
notion of decomposable encryption. However, this approach
significantly \emph{restricts the length of  plaintexts} (e.g., to 3 or 4 bits) to be encrypted, which does not lead to practical solutions to the problem. 
We have provided a detailed evaluation and analysis of existing CB-PWS solutions (see Section~\ref{subsec:performance-anal}).

To our knowledge, the known CB-PWS constructions either require $O(n)$ rounds,
where $n$ is the number of users~\cite{CVH09,LW10,RCV11} or,
by significantly restricting the length of  messages to be encrypted, do not lead to practical solutions.~\cite{KK12} Moreover, the solution in~\cite{KK12}  requires Web search
engines to implement and run the protocols. However, search-engine service providers have no incentive to implement costly protocols that they cannot profit from. 
As yet, there are no known constructions of practical constant-round
CB-PWS systems.

\section{Summary and Ongoing Work}\label{sec:conc}

Web searches have been shown to be sensitive in many cases. Any information leaked from search histories
can endanger user privacy. Search histories may contain
health-related data and  other personal information, including, but
not restricted to, political or religious views and sexual orientation
data.
For example, Google provides signed-in users with personalized search
results based on their search and navigation histories. Furthermore, users typing search queries in the Web interface are
prompted with suggestions derived from their search history. To this end, Google tracks all Web searches performed
by a signed-in user as well as the target Web pages clicked from the
search result page.

In this work, we present a constant-round CB-PWS protocol for
protecting users' privacy when a broadcast channel is available. 
Our solution can be deployed easily in
current systems because it does not require any changes on the service-provider side. However, further work is required:
\begin{itemize}
\item We are attempting to provide a full implementation of our solution
	to demonstrate its practicability.
\item We should improve the performance of the group-manager aspect,
  particularly when the group membership is dynamic. 
\end{itemize}




\end{document}